\documentclass[cits,hyper]{JINST}

\usepackage{amsmath}
\usepackage{amssymb}
\usepackage{hepunits}
\usepackage{hepparticles}
\usepackage{graphicx}
\usepackage[dvips]{epsfig}

\newcommand{\Lumi}{luminosity }

\title{Relative luminosity measurement of the LHC with the ATLAS forward calorimeter}

\author{
\footnotesize{
{\centering The HiLum ATLAS Endcap Collaboration\\[1em]}
A.~Afonin$^8$,
 A.V.~Akimov$^5$,
 T.~Barillari$^6$,
 V.~Bezzubov$^8$,
 M.~Blagov$^5$, 
 H.M.~Braun$^{11}$,
 D.~Bruncko$^3$,
 S.V.~Chekulaev$^{10}$,
 A.~Cheplakov$^2$,
 R.~Degele$^4$,
 S.P.~Denisov$^8$, 
 V.~Drobin$^2$,
 P.~Eckstein$^1$,
 V.~Ershov$^2$,
 V.N.~Evdokimov$^8$, 
 J.~Ferencei$^3$,
 V.~Fimushkin$^2$,
 A.~Fischer$^6$,
 H.~Futterschneider$^1$,
 V.~Garkusha$^8$,
 A.~Glatte$^1$, 
 C.~Handel$^4$,
 J.~Huber$^6$,
 N.~Javadov$^2$,
 M.~Kazarinov$^2$,
 A.~Khoroshilov$^7$,
 A.E.~Kiryunin$^6$,
 E.~Kladiva$^3$,
 M.~Kobel$^{1}$,
 A.A.~Komar$^5$, 
 M.~Komogorov$^2$,
 A.~Kozelov$^8$, 
 G.~Krupny$^8$, 
 V.~Kukhtin$^2$,
 S.~Kulikov$^8$, 
 L.L.~Kurchaninov$^{10}$,
 E.~Ladygin$^2$,
 A.B.~Lazarev$^2$,
 A.~Levin$^8$,
 W.F.~Mader$^1$,
 A.L.~Maslennikov$^7$,
 S.~Menke$^6$,
 L.~Merkulov$^2$,
 A.~Neganov$^2$,
 H.~Oberlack$^6$,
 C.J.~Oram$^{10}$,
 R.~Othegraven$^4$,
 S.V.~Peleganchuk$^7$,
 V.~Petrov$^2$,
 S.~Pivovarov$^7$,
 G.E.~Pospelov$^6$,
 N.~Prokopenko$^8$, 
 J.~Rascvetalov$^8$,
 N.Rusakovich$^2$,
 J.P.~Rutherfoord$^9$,
 D.~Salihagic$^6$,
 A.Y.~Savine$^9$,
 P.~Schacht$^6$,
 H.~Secker$^4$,
 F.~Seifert$^{1}$\thanks{Corresponding author.}~,
 V.~Seleznev$^8$,
 L.~Shaver$^9$,
 S.~Shilov$^2$,
 A.A.~Snesarev$^5$, 
 M.~Soldatov$^8$,
 J.~Spalek$^3$,
 M.~Speransky$^5$, 
 D.~Stoyanova$^8$,
 A.~Straessner$^1$,
 P.~Strizenec$^3$,
 V.V.~Sulin$^5$, 
 A.~Talyshev$^7$,
 S.~Tapprogge$^4$,
 Yu.A.~Tikhonov$^7$,
 Y.~Usov$^2$,
 V.~Vadeev$^2$,
 I.~Vasiliev$^8$,
 R.~Walker$^9$ and
 C.~Zeitnitz$^{11}$\\
\footnotesize{
\llap{$^1$}Technische Universit\"at Dresden, Dresden, Germany\\
\llap{$^2$}Joint Institute for Nuclear Research, Dubna, Russia\\
\llap{$^3$}Institute of Experimental Physics of the Slovak Academy of Sciences, Kosice, Slovakia\\
\llap{$^4$}University of Mainz, Mainz, Germany\\
\llap{$^5$}LPI,  Lebedev Physical Institute,  Moscow, Russia\\
\llap{$^6$}Max-Planck-Institut f\"ur Physik, Munich, Germany\\
\llap{$^7$}BINP, Budker Institute of Nuclear Physics, Novosibirsk, Russia\\
\llap{$^8$}IHEP, Institute for High Energy Physics, Protvino, Russia\\
\llap{$^{9}$}University of Arizona, Tucson, USA\\
\llap{$^{10}$}TRIUMF, Vancouver, Canada\\
\llap{$^{11}$}University of Wuppertal, Wuppertal, Germany\\
}
E-mail: \email{Frank.Seifert@cern.ch}
}}

\abstract{In this paper it is shown that a measurement of the relative luminosity changes at the LHC may be obtained by analysing
the currents drawn from the high voltage power supplies of the electromagnetic section
of the forward calorimeter of the \mbox{ATLAS} detector. The method
was verified with a reproduction of a small section of the ATLAS forward
calorimeter using \mbox{proton} beams of known beam energies and variable intensities at the U-70 accelerator at IHEP in Protvino, Russia. The experimental setup and the data taking during a test beam
run in April 2008 are described in detail. A comparison of the
measured high \mbox{voltage} currents with reference measurements from beam
intensity monitors shows a linear dependence on
the beam intensity. The non-linearities are measured to be less than
\unit{0.5}{\%} combining statistical and systematic uncertainties.}

\keywords{Large detector systems for particle and astroparticle physics; Noble-liquid detectors}

\clearpage

\begin{document}

\section{Introduction}
ATLAS~\cite{atlas} is a multi-purpose physics detector
at the Large Hadron Collider (LHC)~\cite{LHC} at CERN. It records the
products of proton-proton collisions at centre-of-mass energies up to
14~\TeV. The instantaneous luminosity of the LHC is planned to be \unit{$10^{34}$}{\lumiunits}. For many of the anticipated
physics analyses, in particular for measurements of absolute
cross-sections, the knowledge of the integrated luminosity is
essential. It will be provided by
the luminosity detectors LUCID~\cite{lumi-tdr,atlas}, used for a
relative luminosity measurement, and ALFA~\cite{lumi-tdr,alfa}, used for an \mbox{absolute} calibration of the luminosity determination at low instantaneous luminosities of about \unit{$10^{27}$}{\lumiunits}. For the absolute calibration, a precision of about 3\% is aspired~\cite{lumi-tdr}. For the relative measurement using LUCID, a statistical uncertainty of about 1\% is expected after 3 minutes of recording time~\cite{lum-pub-2006-001}. The information about relative changes in luminosity are also important for
monitoring beam stability and beam degradation in order to efficiently
operate the ATLAS trigger and data acquisition system.\\
\clearpage

Here, we discuss an alternative method to obtain a continuous
measurement of the relative change in instantaneous luminosity at the
LHC using the currents drawn from the high voltage (HV) power supplies
in the electromagnetic section of the ATLAS liquid-argon forward
calorimeter, FCal1~\cite{atlas,Bonivento:684140}.\\

It is based on the measurement of the currents induced by ionisation of the liquid argon (LAr) by particles produced in proton-proton collisions. The dominant physics process at the LHC is inelastic proton-proton scattering, with a cross-section of about \unit{$80$}{$mb$} \cite{Pythia,Phojet}. For each bunch crossing an average of 23 inelastic events \cite{Pythia}
will be produced at nominal LHC luminosity. They deposit most of their
energy in the forward section of the detector. The ionisation and thus
the HV currents are expected to scale linearly with the instantaneous
luminosity up to the nominal LHC luminosity. While \mbox{LUCID} can in principle provide a measurement for every proton bunch, the method presented here is limited by the HV current readout cycle time of several seconds. Therefore only a determination averaged over many bunch-crossings will be possible with this method. The feasibility of this method is verified in a test beam at the U-70 proton
accelerator~\cite{U70} in Protvino, Russia, using a test
module of the FCal1. Possible non-linearities could be caused by
a reduction of the HV at high currents or positive ion build-up in the
gaps~\cite{pib}.\\

This paper is organised as follows: In Section 2 the beam parameters
and beam structure of the U-70 accelerator in Protvino are described,
as well as the experimental setup and the beam monitoring. Furthermore, the test module of the FCal1 and the readout
of the currents drawn from the HV power supplies are described.
Section 3 discusses the analysis of the data. A summary and conclusion
are given in Section 4.

\section{Test beam configuration}
\subsection{Purpose of the test beam run}

The aim of the HiLum project~\cite{Hilum-Nim,Peter-Como,Yuriy-Pisa} is the study of the
behaviour of the electromagnetic endcap calorimeter, EMEC~\cite{atlas}, the hadronic endcap calorimeter, HEC~\cite{atlas},
and the electromagnetic section of the forward calorimeter, FCal1, in
the environment of high particle rates expected for the LHC upgrade
phase (sLHC)~\cite{sLHC}. At the sLHC instantaneous luminosities of up to \unit{$10^{35}$}{\lumiunits} are foreseen. The high interaction rate
and particle flux in the test beam were obtained by placing small calorimeter test
modules directly into the Protvino accelerator proton beam behind iron
absorbers. The layout and material of these modules are similar to that used in the ATLAS detector. Different LHC luminosities have been simulated experimentally by varying the beam intensities at a constant beam energy of \unit{60}{\GeV} using the bent crystal technique~\cite{bentcrystals} for beam extraction. The expected energy flux through the calorimeters was calculated to correspond to operating conditions in \mbox{ATLAS} and the test beam setup was optimised accordingly by Monte-Carlo simulations~\cite{Hilum-Nim}.\\

During the test beam periods the U-70 was filled with five bunches of
protons ($p$) separated by a gap of about \unit{$990$}{\ns}. Each bunch was of \unit{$166$}{\ns} length, but only about \unit{$30$}{\ns} were occupied by protons. One accelerator fill was extracted over a time interval of about
\unit{$1.2$}{\second}, henceforth referred to as `spill length'. The spill cycle time was about \unit{$9.5$}{\second} leading to a gap
of about \unit{$8.3$}{\second} between two spills.  Beam
intensities were ranging from \unit{$\power{10}{7}$}{p/$spill$} to \unit{$\power{10}{12}$}{p/$spill$} after extraction.

\subsection{Experimental setup and beam monitoring}

\begin{figure}[b]
\begin{center}
\includegraphics[width=0.99\textwidth] {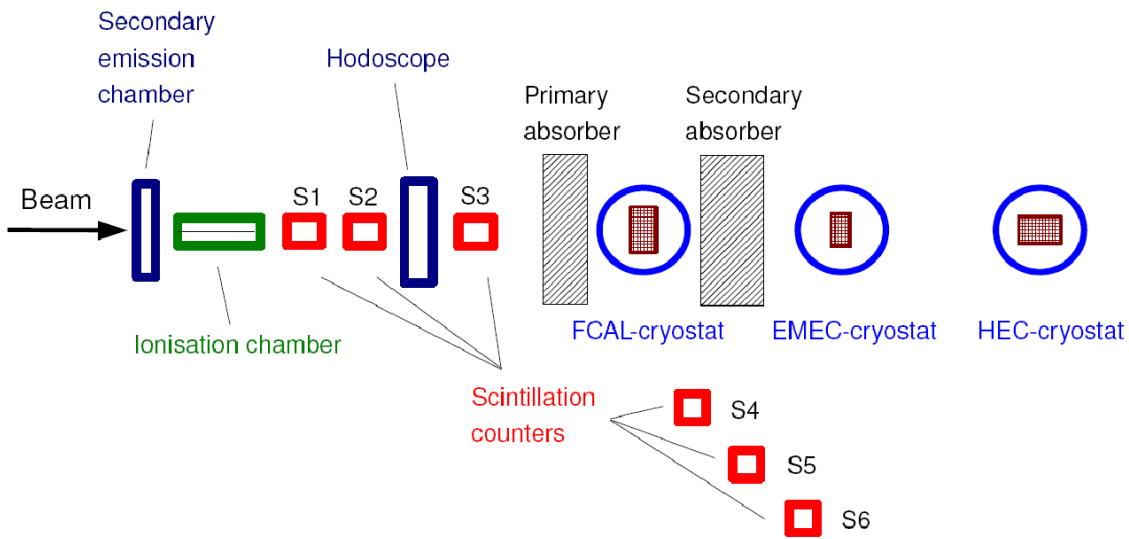}
\caption[Experimental setup in Protvino.]
{Schematic layout of the experimental setup of the high luminosity
test beam. The beam direction is from left to right. See text for details.}
\label{setup}
\end{center}
\end{figure}

The experimental setup used at the high luminosity beam line in
Protvino is shown in figure~\ref{setup}.  Several beam monitoring
devices were installed between the extraction point and the
calorimeter cryostats.  An evacuated secondary emission chamber containing a matrix of \unit{5}{\milli\meter} wide electrode strips was used for beam position measurements in the high intensity range. An ionisation chamber provided accurate measurements of the integrated beam intensity per spill up to \unit{$\power{10}{11}$}{p/$spill$}. Its absolute calibration was done using activated aluminium foils. A scintillation counter hodoscope for beam profile measurements and six scintillation counters, three in the beam line (S1-S3) and three at large angles (S4-S6) with respect to the beam, were also installed. They were used as a cross-check for the ionisation chamber
measurements and served as a bunch trigger as well. The hodoscope and the counters S1, S2, and S3 were operational only at low intensities up to \unit{$5\cdot\power{10}{7}$}{p/$spill$} and had to be moved out of the beam at higher intensities to avoid
damage. The counters S4, S5, and S6 were detecting secondary particles
from beam interactions with the absorber material and could therefore be used during high intensities.

\subsection{The FCALchick and the readout of the HV currents}

\begin{figure}[t]
\begin{center}
\includegraphics[width=0.495\textwidth] {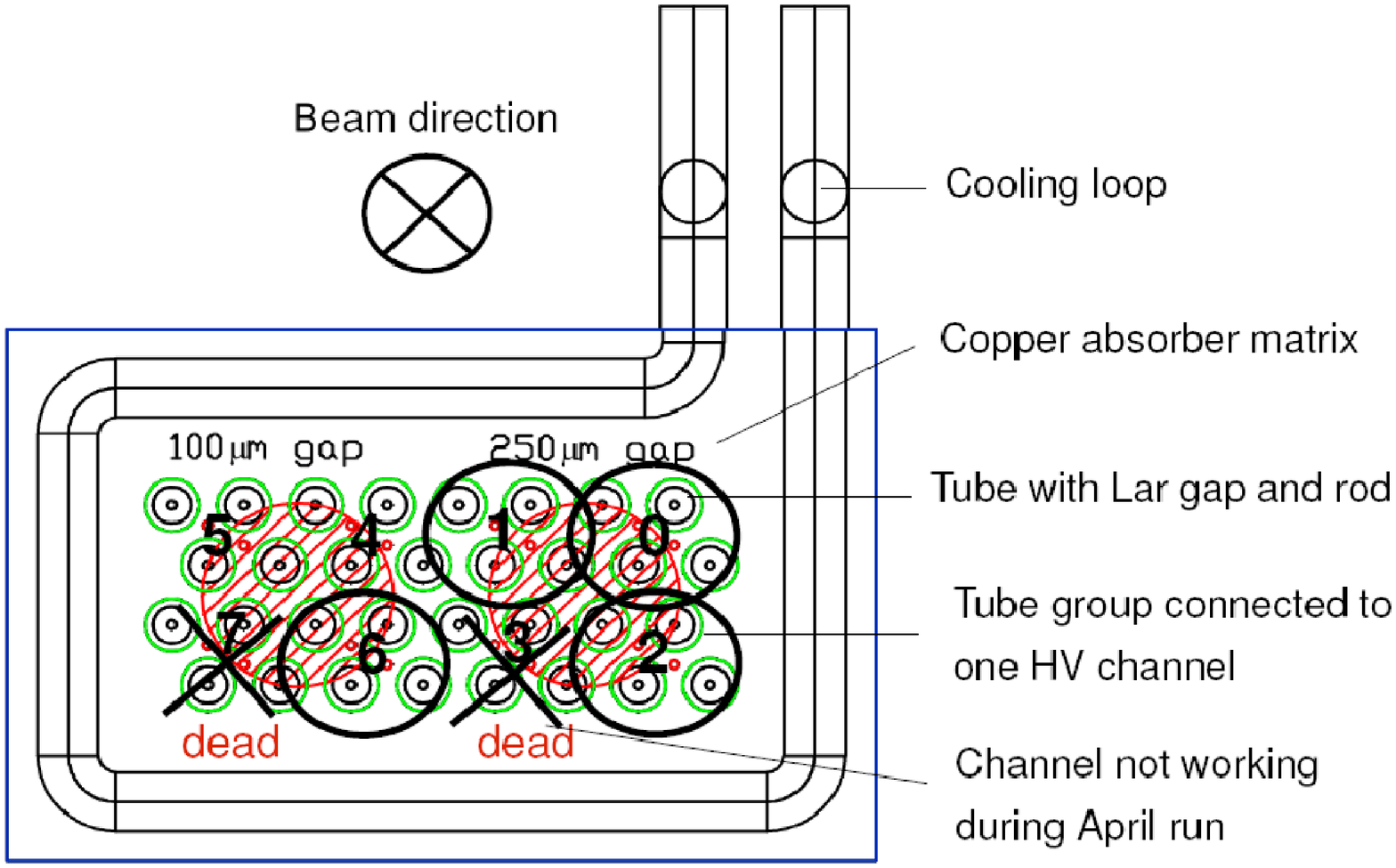}
\includegraphics[width=0.495\textwidth] {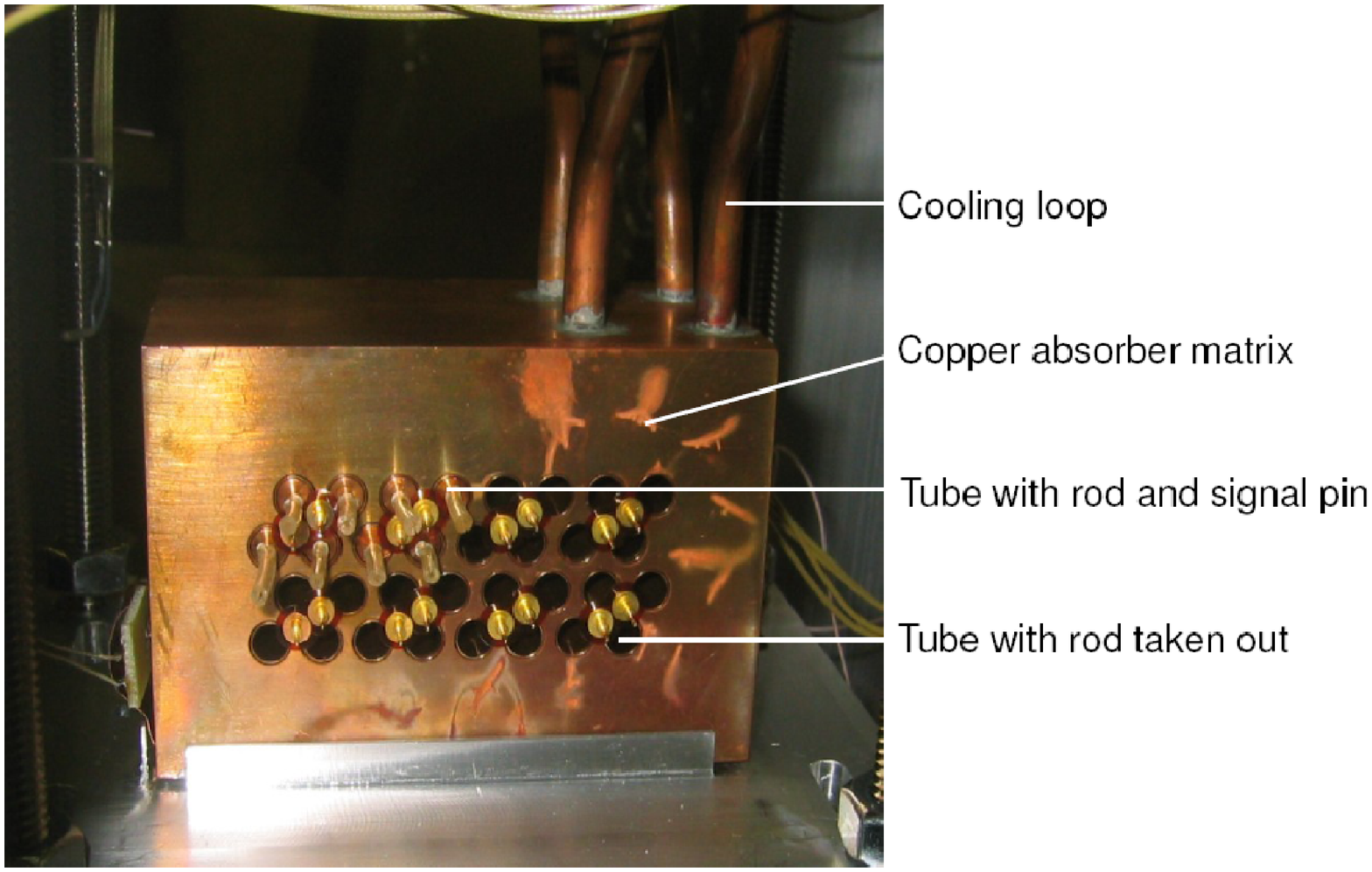} 
\caption[The FCALchick.]{Left: Channel layout during the April
2008 data taking run. Channels surrounded by circles were connected to the additional measurement device. Channels crossed out were not
operational during the April run. Right: The FCALchick used in the
Protvino test beams. The copper absorber matrix holds the rods with
signal pins. Some of the rods are not installed. Also visible is the
cooling loop for liquid-nitrogen cooling.}
\label{fcalchick}
\end{center}
\end{figure}

The FCal1 in ATLAS consists of a copper absorber matrix in which a hexagonal array of concentric copper tubes (anodes) and
copper rods (cathodes) with \unit{$250$}{\micron} wide liquid-argon gaps
between is embedded~\cite{atlas,johnspaper}. The prototype of the FCal1 used in
Protvino, referred to as the FCALchick, consisted of one section of 16
electrodes with \unit{$250$}{\micron} LAr gaps and one section of 16 electrodes with \unit{$100$}{\micron} LAr gaps, a prototype of the design being proposed for the sLHC.  The electrodes of the FCALchick are only
\unit{$50$}{\milli\meter} long compared to \unit{$450$}{\milli\meter}
\cite{atlas,johnspaper} in the ATLAS FCal1. An additional liquid-nitrogen loop is
included for extra cooling.  The layout and a picture of the FCALchick
are shown in figure~\ref{fcalchick}. The shaded circles indicate the
beam size in front of the absorbers. Groups of four
electrodes were connected in parallel. The eight groups of the FCALchick
were connected to a signal readout wire and a high voltage supply
wire. The low pass filter for each HV channel had resistors
of \unit{$10$}{\kilo\ohm} and filter capacitors of
\unit{$220$}{\nano\farad}, corresponding to a time constant of
\unit{$2.2$}{\milli\second}.\\

\begin{table}[b]
\centering
\caption[FCALchick operating parameters.]{Beam intensities and
FCALchick (\unit{250}{\micro\meter} side) HV currents estimated before
the test beam run for the various beam intensities. 
The parameters are extrapolated linearly to the highest intensities.\\}
\begin{tabular}{|lllllll|}
\hline
Protons/spill                          & $\power{10}{7}$  & $\power{10}{8}$  & $\power{10}{9}$  & $\power{10}{10}$  & $\power{10}{11}$     & $\power{10}{12}$ \\
Protons/bunch                          & $5$              & $50$             & $500$            & $5000$            &$5\cdot\power{10}{4}$ & $5\cdot\power{10}{5}$\\
LHC luminosity equivalent [\lumiunits] & $\power{10}{32}$ & $\power{10}{33}$ & $\power{10}{34}$ & $\power{10}{35}$  & $\power{10}{36}$     & $\power{10}{37}$ \\
HV current\,/\,channel [\micro\ampere] & $0.12$           & $1.2$            & $12$             & $120$             & $1200$               & $12000$          \\
\hline
\end{tabular}
\label{johnsestimations}
\end{table}

The expected HV currents drawn by the FCALchick readout channels for different beam intensities are shown in table \ref{johnsestimations}, together with the corresponding LHC luminosities. The comparison to the various luminosities in ATLAS which gives the same HV current density as observed in the FCalchick is obtained by using the GEANT4~\cite{Geant4} simulation code of this test beam setup along with GEANT4 simulations of minimum bias proton-proton collisions in the \mbox{ATLAS} detector. Both simulations give the rate of ionisation in the \unit{$250$}{\micron} FCal-style gaps. For the \mbox{ATLAS} simulation comparisons, gaps near the highest pseudorapidity covered by the forward calorimeter were chosen at a depth in the calorimeter near the electromagnetic shower maximum.\\

In the test beam setup an ISEG~\cite{ISEG} HV supply module with effective 16\,bit analog-to-digital converters (ADC) for internal HV current measurement was installed, leading to a
resolution of about \unit{$200$}{\nano\ampere} over a range of
\unit{$\pm10$}{\milli\ampere}. The readout cycle time was
\unit{$1$}{\second}. This was estimated to be insufficient for the proposed measurements, especially considering the expected HV currents in the low intensity region (compare table \ref{johnsestimations}).\\ 

\begin{figure}[t]
\begin{center}
\includegraphics[scale=0.5] {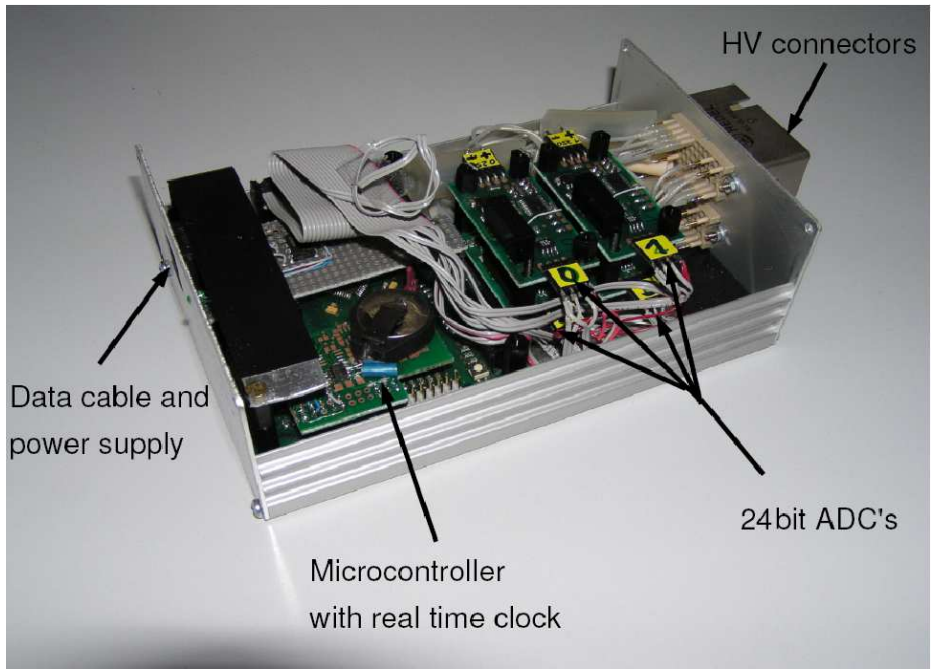}
\caption[Dresden HV current measurement device.]{Device for the
additional independent measurement of the HV current. The four
24\,bit ADCs for the four HV current readout channels of the
FCALchick are connected to a microcontroller to process their data and
to add a timestamp from the real time clock to each measurement. The
readout frequency is \unit{10}{$Hz$} per channel and the
resolution is \unit{$1.2$}{\nano\ampere} per bit.}
\label{box}
\end{center}
\end{figure}

To improve the precision, an external measurement
device for the HV current with 24\,bit ADCs and a resolution of
\unit{$1.2$}{\nano\ampere} per bit was used. The logging rate of the
device was at \unit{10}{$Hz$} per channel and the
precision of the time-stamp was \unit{$10$}{\milli\second}.
Electronic noise mainly from the HV power supply limited
the effective resolution to about \unit{$25$}{\nano\ampere}, depending
on the channel. A picture of the measurement device is shown in figure~\ref{box}. The data transfer and independent power supply of the measurement device
was realised by a LAN-cable connecting it to a data acquisition PC. In ATLAS, the HV power supplies use 20\,bit ADCs for internal HV
current readout providing a sufficient precision without an external
ammeter.

\subsection{Data taking}

During test beam data-taking it was possible to move the cryostats horizontally,
perpendicular to the beam line such that either
the \unit{$250$}{\micron} side or the \unit{$100$}{\micron} side of
the FCALchick could be centered on the beam. During the runs labelled as 230 and 240-244 the position of the FCALchick cryostat was adjusted such that the beam was centered to the \unit{$250$}{\micro\meter} side. The cryostat position was kept constant, and the running conditions were stable.
These runs are thus used for further
analysis. Channel\,3 of the \unit{$250$}{\micron} side of the FCALchick was not functioning during the April run and the corresponding ADC was therefore connected to
channel\,6 of the \unit{$100$}{\micron} side as indicated in
figure~\ref{fcalchick}.\\

As a reference for the beam intensity, the measurements of the
calibrated ionisation chamber were used, providing one integrated
measurement per spill. Periods with different beam positions relative
to the FCALchick were analysed separately, because the particle flux
through the calorimeter, and thus the HV current, depends on the
calorimeter position relative to the beam while the beam intensity
measured by the ionisation chamber is independent of the calorimeter
position.\\

\section{Analysis of the test beam data}
\subsection{Analysis of HV currents}

\begin{table}[t]
\centering
\caption[Thresholds to separate spill data from noise.]
{HV current thresholds for separating spill data from noise in units
of nA. The low intensity run 230 is treated separately from the high
intensity runs 240-244.\\}
\begin{tabular}{|l|cccc|}
\hline
Run     & Channel 0 & Channel 1 & Channel 2 & Channel 6 \\    
\hline
230     & 470       & 100       & 400       & 80        \\ 
240-244 & 600       & 150       & 460       & 70        \\         
\hline
\end{tabular}
\label{threshs}
\end{table}

The HV current was measured every \unit{100}{\milli\second}. However,
the system was not synchronized with the beam trigger and no
additional information was available on whether a measurement took
place within a spill or outside. For this reason it was necessary to
separate the measurements of the HV current during a spill from those
between two spills where dark current background and electronic noise
dominates. This separation was done by requiring that measurements
within a spill have to be at least three standard deviations above the
noise level. All measured currents above the threshold were assigned
to a spill, whereas all currents below the threshold were considered noise. The thresholds applied are summarised in
table~\ref{threshs}.\\

The differences in the thresholds from channel to channel are due to different dark currents and electronic noise in the corresponding channels. The individual channels of the HV power supplies showed different noise levels, whereas the dark currents have their origin in ground loops or leakage currents in the calorimeter. Slight changes of these conditions were also seen from run to run leading to the different thresholds between the low-intensity run 230 and the runs 240-244 with higher intensity.\\

\begin{figure}[b]
\begin{center}
\includegraphics[width=0.7\textwidth] {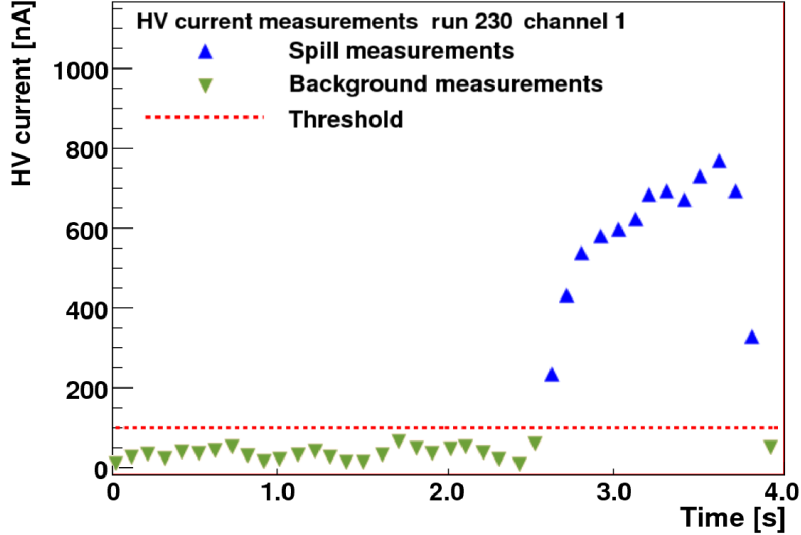} 
\caption[HV current measurements of one spill of run 230.]
{Measurements of the HV currents for one spill taken during
run 230 for one of the four readout channels. Upward pointing
triangles indicate measurements assigned to a spill. Downward pointing
triangles correspond to background measurements between the
spills. The horizontal line shows the threshold to separate the spills
from background.}
\label{spillhistos230}
\end{center}
\end{figure}

The measurements obtained for one channel of one
selected spill in run 230 are plotted in figure~\ref{spillhistos230}. For the data analysis the mean of the background measurements between
the previous and the current spill was subtracted from each measurement
within the spill.\\
\clearpage

The integrated HV current for every spill was
calculated for each channel independently as:

\begin{equation}
I=\sum_{i=1}^{n}(s_i-B)\cdot d.
\label{integral}
\end{equation} 

Here, $s_i$ are the measurements within the given spill, $n$ is the
number of such measurements, $B$ is
the mean of the background measured before the given spill and $d$ is the time interval
between two measurements within the spill. This integral is hence used for
the comparison of the HV currents with the beam intensity measured by the ionisation chamber spill by spill.\\

The main contribution to the uncertainty on the integral is due to the electronics noise. It could hardly be calculated from the fluctuations within the spills because beam intensity variations would cause an additional contribution. Therefore this uncertainty was estimated from the fluctuations of the background measurements. Using the uncertainty on the mean of the background, $\triangle B$, with

\begin{equation}
\triangle B = \frac{1}{\sqrt{m\cdot(m-1)}}\sqrt{\sum_{i=1}^{m}(B-b_i)^2},
\label{averagebackerror}
\end{equation}

the uncertainty on the integrated beam intensity is then given by:

\begin{equation}
\triangle I=\sqrt{[(n\cdot d\cdot\triangle B)^2 + n\cdot (d\cdot\triangle s_i)^2]} =
n\cdot d\cdot \triangle B \cdot \sqrt{1+\frac{m}{n}},
\label{integralerrorsimple}
\end{equation}
\\[5em]
where $b_i$ are the background measurements before the given spill and $m$ is the number of such measurements. The uncertainty on a
single measurement, $\triangle s_i$, is estimated from the data in the
time interval between spills as $\triangle s_i = \sqrt{m}\cdot
\triangle B$.\\

The spills identified in the HV data stream were
matched to the ionisation chamber measurements using a
\unit{3}{\second} time window, much smaller than the
\unit{9.5}{\second} interval between spills.  The integral calculated
from Eq.~\ref{integral} with the corresponding uncertainty
calculated from Eq.~\ref{integralerrorsimple} are then compared
to the beam intensity measurements of the ionisation chamber.\\

Very few cases were found where the beam intensity measured by the ionisation
chamber was inconsistent with those from the scintillation
counters, possibly due to readout instabilities. These data were removed from the subsequent analysis.

\subsection{Beam position variations}

\begin{figure}[b]
\begin{center}
\includegraphics[width=0.495\textwidth] {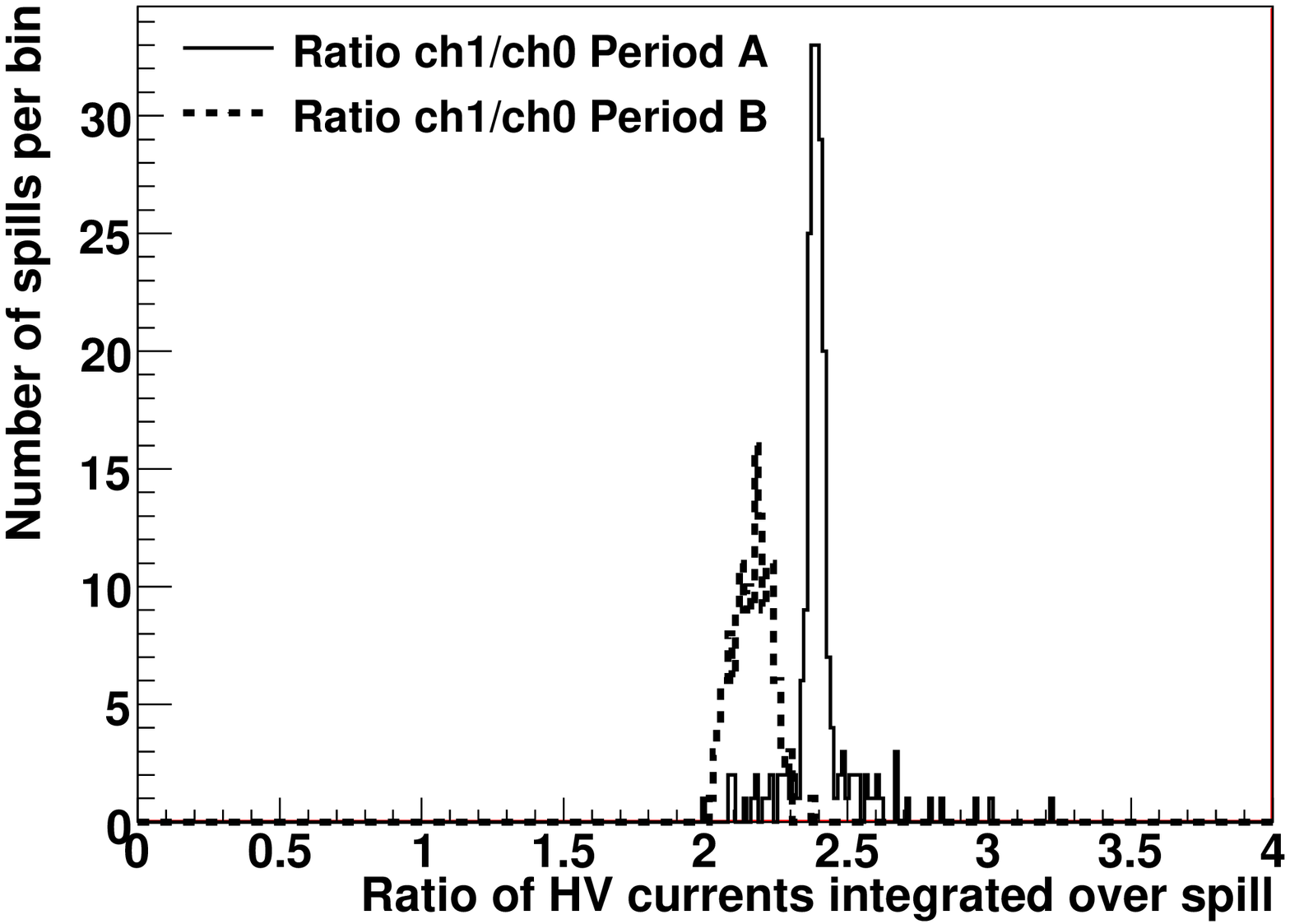}
\includegraphics[width=0.495\textwidth] {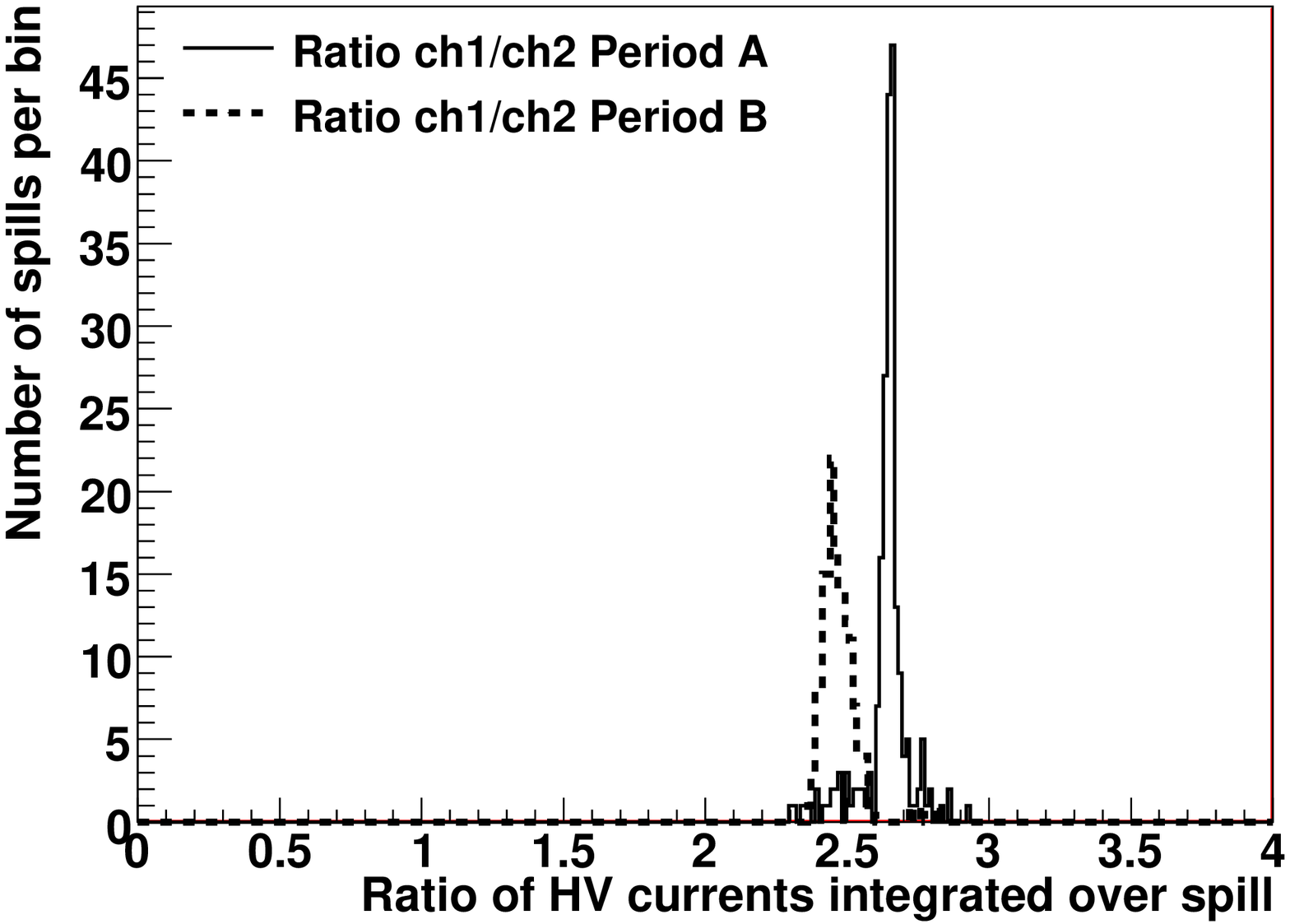}
\caption[Ratios of HV currents for runs 230, 240-244.]{Ratios of
HV currents between channel\,1 and channel\,0 in the left plot and
channel\,1 and channel\,2 in the right plot for runs 230, 240-244. For
each channel the integrated HV current over one spill is used to
calculate the ratios. \\}
\label{ratios230-244}
\end{center}
\end{figure}

The position of the FCALchick cryostat was kept constant during the runs 230 and 240-244, but small \mbox{variations} of the beam impact point could not be excluded. To investigate these variations, the integrated
currents of the four channels were compared. A
significant change of the ratio of currents \mbox{between} two
channels would indicate a shift of the beam position. All six possible ratios were analysed: channels\,0-1 and 2-6 for
horizontal variations, channels\,0-2 and 1-6 for vertical variations
and channels\,0-6 and 1-2 for the diagonal directions (see the
channel layout in figure~\ref{fcalchick}). Two ratios \mbox{between} the
different channels are shown in figure~\ref{ratios230-244}, indicating a
horizontal displacement in the left plot and a diagonal displacement
in the right plot.\\

Two clearly separated peaks can be identified. Each of the two peaks corresponds to one of two run periods, period A
containing runs 230, 240, and 241 and period B containing runs 242-244,
with nearly constant beam position within each period. Because the
exact profile of the proton beam after the primary absorber is
poorly known, the difference in the HV current of the four channels for
different beam positions cannot be easily predicted. Therefore the
two data taking periods were analysed separately.

\subsection{Investigation of non-linearities}

To investigate possible non-linearities between the HV current and the
beam intensity, a second order polynomial of the form

\begin{equation}
I=P_{1}\cdot J + P_{2}\cdot J^{2},
\label{nonlinearrelation}
\end{equation}

was used to fit the data. Here, $I$ is the HV current integrated over the spill in units of nC, and $J$ is the beam
intensity in units of $10^{6}p$/spill, as measured by the ionisation chamber with the linear coefficient $P_{1}$, and the coefficient of the
quadratic term, $P_{2}$. The non-linear fraction, $N$, of the HV current was calculated as:

\begin{equation}
N(J)=\frac{P_{2}}{P_{1}}\cdot J,
\label{nonlinearfraction}
\end{equation}

depending linearly on the beam intensity $J$.\\ 

\begin{figure}[t]
\begin{center}
\includegraphics[width=0.65\textwidth] {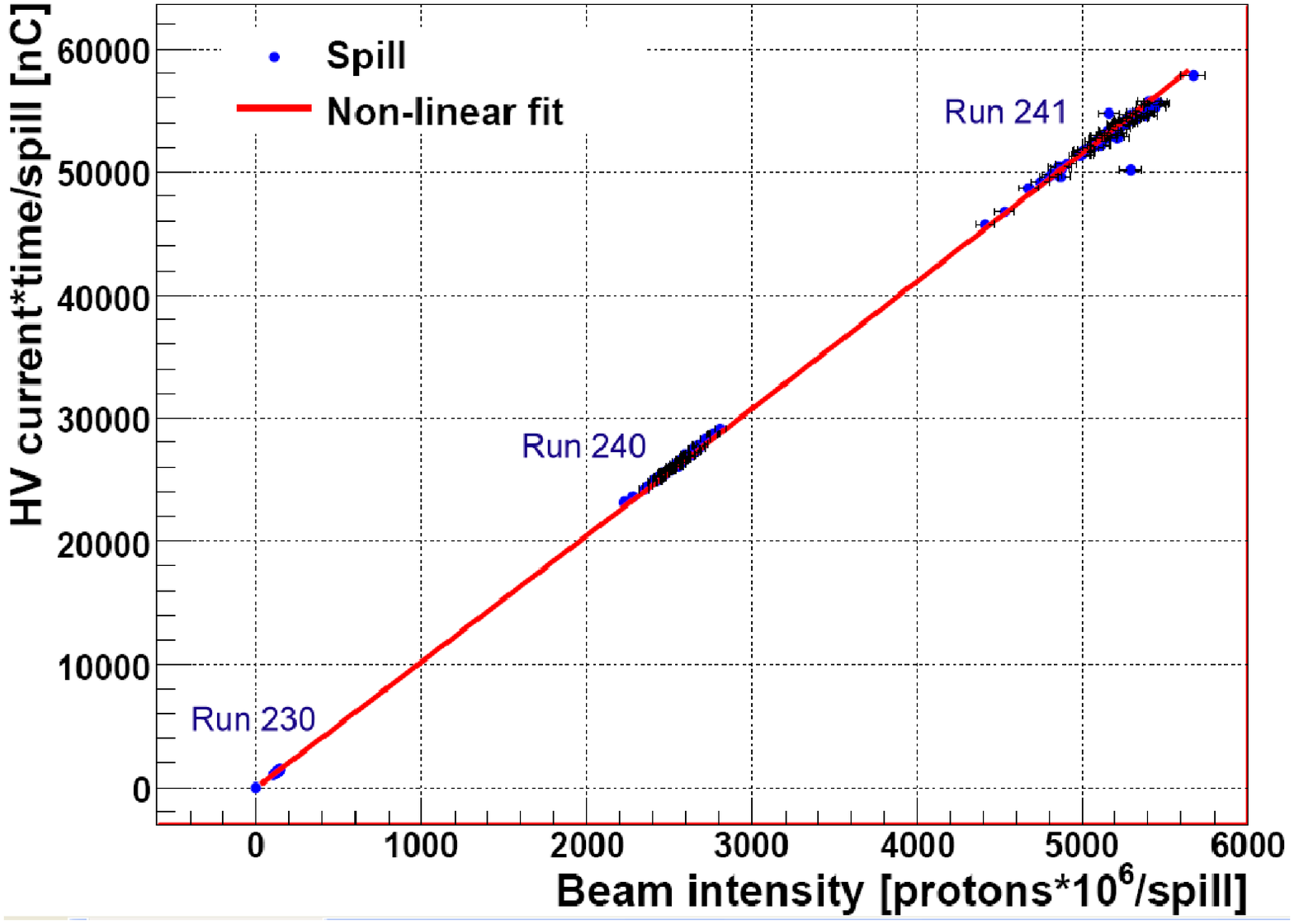}
\includegraphics[width=0.65\textwidth] {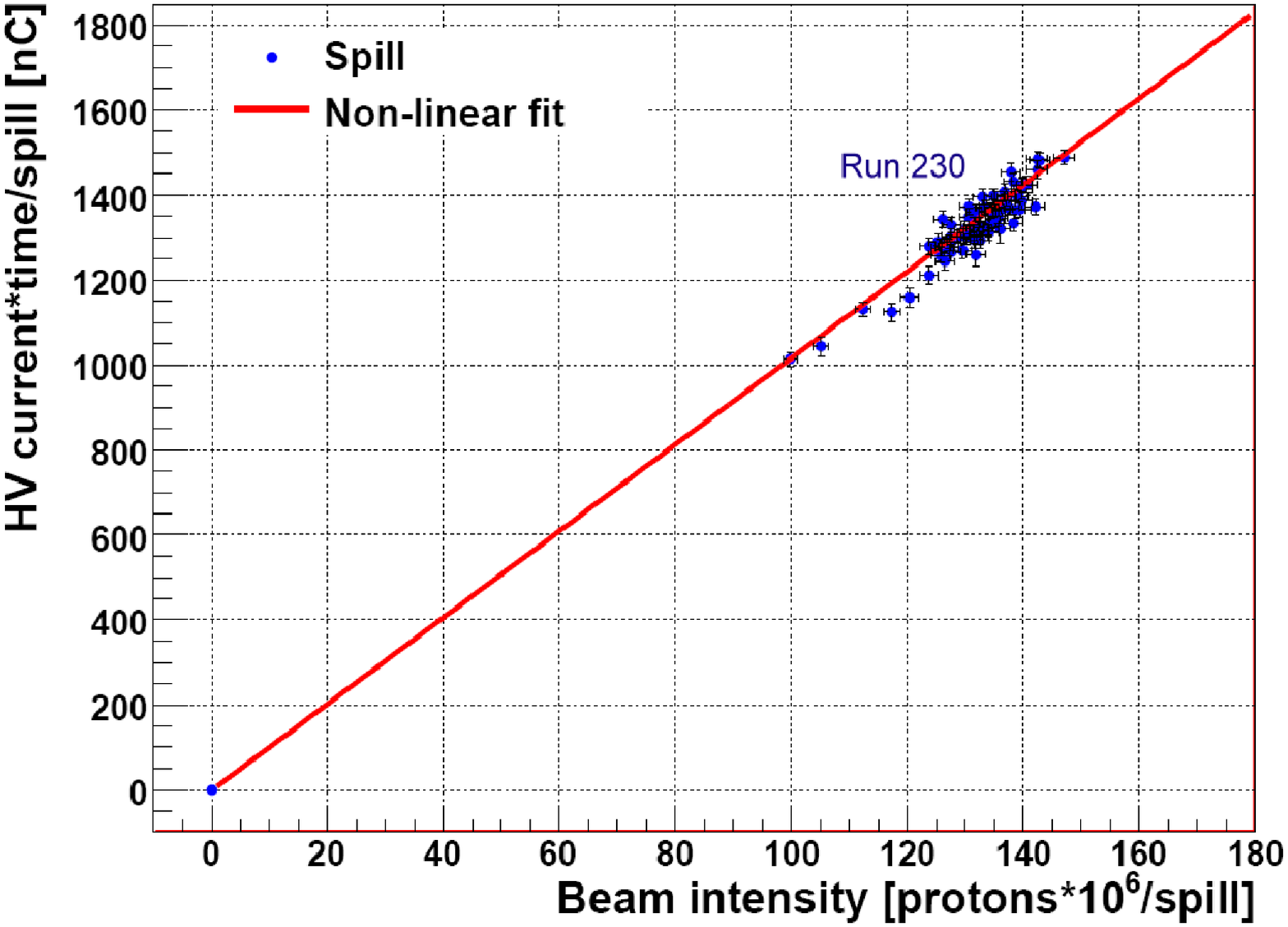}
\includegraphics[width=0.65\textwidth] {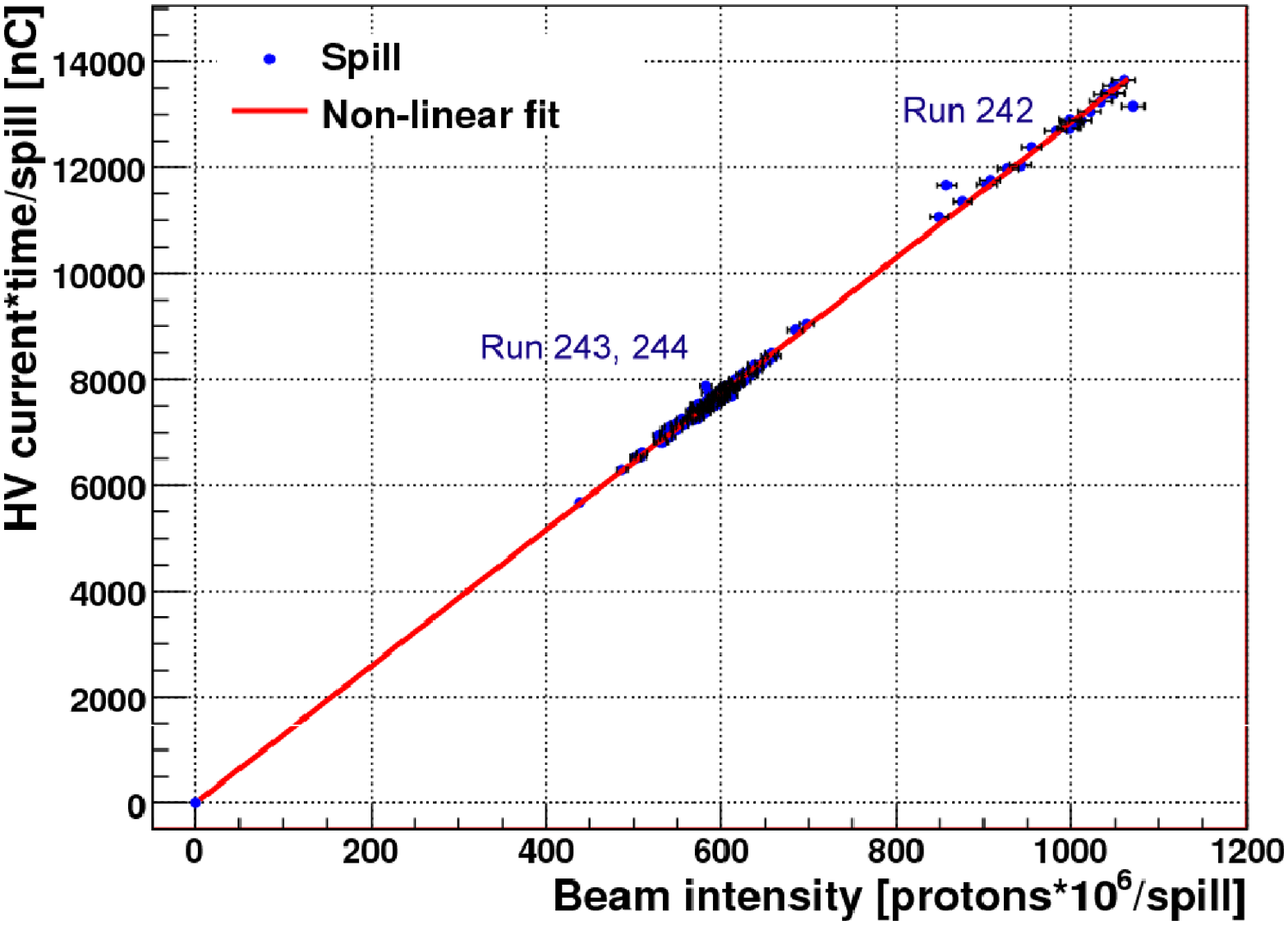}
\caption[HV current summed vs. beam intensity for runs 230, 240, 241 with non-linear fit.]{Measured HV current summed over the channels 0, 1, and 2 vs. beam intensity compared to a non-linear fit for period A (runs 230, 240, 241) in the top plot, a magnified view of run 230 in the middle plot, and period B (runs 242-244) in the bottom plot.} \label{resultfirstsum}
\end{center}
\end{figure}

A comparison of the HV current to the beam intensity measured by the ionisation chamber, both integrated over each spill, is shown in figure~\ref{resultfirstsum} together with the results of the fit for the two periods A and B. Also given in the middle plot of figure~\ref{resultfirstsum} is a magnified view of run 230 at lowest intensities. The fit is a combined fit of the channels 0, 1 and 2 which means it is applied to the HV current summed over these three channels in dependence on the beam intensity. These are the channels with \unit{$250$}{\micro\meter} LAr gaps, whereas the fourth ADC was connected to channel\,6 of the \unit{$100$}{\micro\meter} side.\\

The parameters obtained in these fits for the two periods are
summarised in table~\ref{fitparslowerror}. The fit parameters in the
lines labelled as ''$\sum(0,1,2)$'' were taken from the combined fit. From the calibration with
activated aluminium foils it can be concluded that the relative
precision of the beam intensity measurement by the ionisation chamber
is better than 5\% which is the accuracy of the calibration
method. This is however only an upper bound on the
uncertainty. Assigning a reduced uncertainty of 1.2\% to the measurements
of the ionisation chamber a much better behaviour of the $\chi^{2}$
per degree of freedom of the fits was found. This value was used for obtaining the final result. The quadratic contribution of the term is given for
$J=\unit{$10^9$}{p/$spill$}$ which corresponds to the nominal LHC
\Lumi of \unit{$10^{34}$}{\lumiunits} (see table~\ref{johnsestimations}).\\ 

Using the summed current, the measured non-linear fraction is
\unit{$(0.26\pm0.05)$}{\%} for period A and \unit{$(0.59\pm0.65)$}{\%}
for period B at a beam intensity of \unit{$10^9$}{p/$spill$}. The
reason for the larger uncertainty on the non-linear fraction in period B is
due to the smaller beam intensity range up to
\unit{$1.1\cdot10^9$}{p/$spill$} compared to a maximum of
\unit{$5.5\cdot10^9$}{p/$spill$} for period A.

\clearpage

\begin{table}[t]
\centering
\caption[Fit parameters for the runs 230 and 240-244.]{Fit
parameters of the non-linear fit for period A (runs 230, 240, 241) and
period B (runs 242, 243, 244). The unit $10^{6}p$/spill is used for the beam intensity to obtain the fit parameters. A scaling factor of $10^{3}$ is applied
in the last column in order to obtain the non-linear fraction at a beam
intensity of $10^{9}p/$spill which corresponds to the nominal
LHC \Lumi.\\}

\begin{tabular}{| ll | r@{$\pm$}l r@{$\pm$}l c r@{$\pm$}l r@{$\pm$}l |}
\hline
Period & Channel    & \multicolumn{2}{c}{$P_1$} & \multicolumn{2}{c}{$P_2\cdot\power{10}{5}$} & $\chi^2($DoF$)$ & \multicolumn{2}{c}{$P_2/P_1\cdot\power{10}{6}$}	& \multicolumn{2}{c|}{$N(J)\cdot\power{10}{2}$} \\
& & \multicolumn{2}{c}{ } & \multicolumn{2}{c}{ } & & \multicolumn{2}{c}{ } & \multicolumn{2}{c|}{$J=10^{9}~p/$spill$$} \\
\hline
A      & $0$        		& $2.39$&$0.01$  &  $0.08$&$0.17$  					 & $326(217)$       & $0.35$&$0.70$  							   & $0.03$&$0.07$ \\  
       & $1$         		& $5.64$&$0.01$  &  $2.11$&$0.25$  					 & $277(217)$       & $3.74$&$0.45$  							   & $0.37$&$0.05$ \\
       & $2$         		& $2.17$&$0.01$  &  $-0.14$&$0.14$ 					 & $225(217)$       & $-0.63$&$0.64$ 							   & $0.06$&$0.06$ \\
       & $6$         		& $3.29$&$0.01$  &  $-0.53$&$0.16$ 					 & $186(217)$       & $-1.60$&$0.48$ 							   & $0.16$&$0.05$  \\
       & $\sum(0,1,2)$	& $10.17$&$0.02$ &  $2.69$&$0.51$  					 & $241(217)$     	& $2.64$&$0.51$  							   & $\textbf{0.26}$&$\textbf{0.05}$	\\
\hline
B      & $0$         		& $3.31$&$0.01$  & $-19.0$&$2.1$  					 & $457(161)$       & $-57.4$&$6.6$ 							   & $5.74$&$0.66$  \\
       & $1$         		& $6.69$&$0.02$  & $32.1$&$4.6$   					 & $189(161)$       & $48.0$&$7.1$  							   & $4.80$&$0.71$  \\
       & $2$         		& $2.89$&$0.01$  & $-15.4$&$1.8$  					 & $204(161)$       & $-53.2$&$6.3$ 							   & $5.32$&$0.63$  \\
       & $6$         		& $2.14$&$0.01$  & $12.9$&$1.6$   					 & $206(161)$       & $60.3$&$7.6$  							   & $6.03$&$0.76$  \\
       & $\sum(0,1,2)$	& $12.93$&$0.05$ & $-7.6$&$8.1$   					 & $126(161)$       & $-5.9$&$6.2$ 			  				   & $\textbf{0.59}$&$\textbf{0.62}$ \\
\hline
\end{tabular}
\label{fitparslowerror}
\end{table}

\subsection{Comparison of HV currents measured externally with HV power supply currents}

\begin{figure}[t]
\begin{center}
\includegraphics[width=1.0\textwidth] {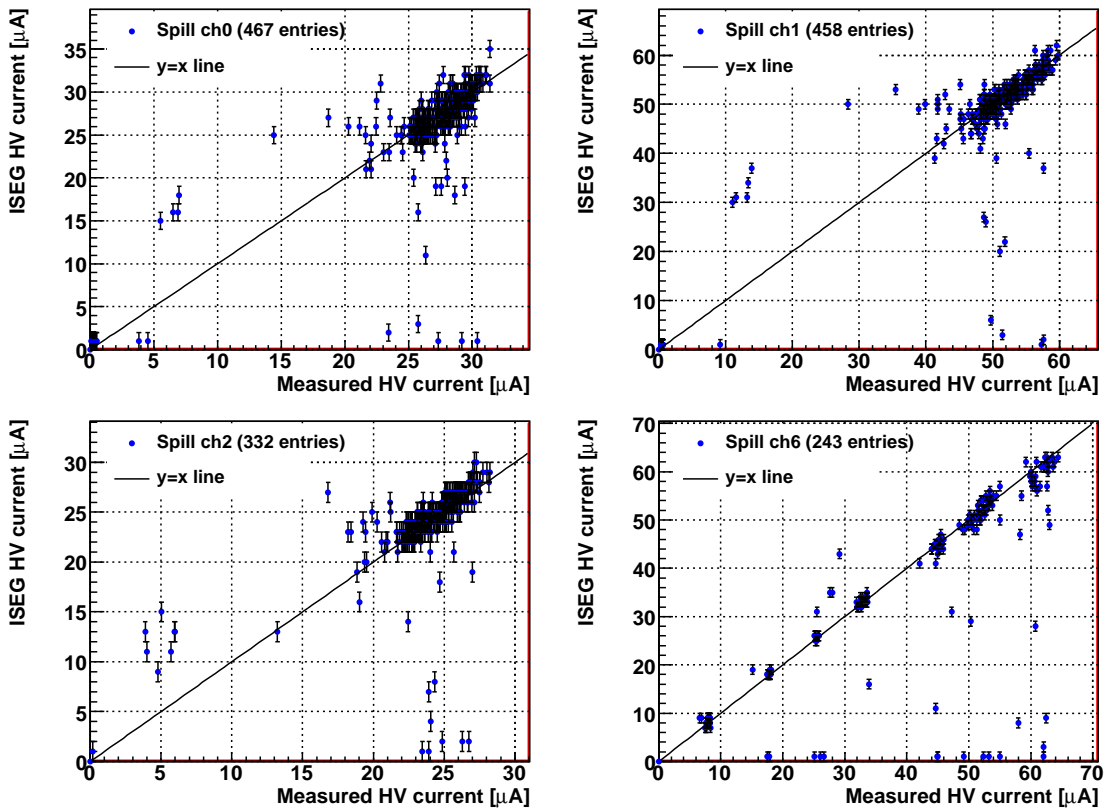}
\caption[ISEG HV current summed vs. Dresden HV current for runs 211-223 with x=y line.]
{Comparison of the ISEG internal measurements of the HV current and the HV
current measured by the external measurement device for runs 211-223
on a spill-by-spill basis.}
\label{PVSS_211-223}
\end{center}
\end{figure}

In ATLAS the measurement of the FCal HV currents will be performed
directly by the ISEG HV power supplies~\cite{johnspaper}. To ensure that this method of measuring the relative \Lumi can be used in ATLAS
it has to be shown that the HV currents measured by the external measurement device and the ISEG internal
measurement device are in agreement. One example for such a comparison is shown in figure~\ref{PVSS_211-223}. The integral of the current over the U-70 accelerator spill as measured by the external device was divided by the spill length. This value was compared to the ISEG measurements in the corresponding spill stored in the PVSS data archive~\cite{PVSS}. As only one single measurement per spill with a precision of the timestamp of \unit{1}{\second} was available for the PVSS data stream, it is unknown where within the spill the measurement took place. The data points which are not in agreement with an ideal linear dependence are therefore due to measurements close to the start or end of the spill, where the intensity is varying strongly.\\

Since the two ammeters have different low pass filters the time
constants to resolve the spill variation are not the same. These
constants are \unit{2.2}{\milli\second} and \unit{48.2}{\milli\second}
for the ISEG module and the external device, respectively. These
deviations from linearity are well reproduced when simulating the
signal responses. When taking the different time constants into
account, the measurements of both devices are in very good
agreement. In conclusion, the ISEG HV module will be suitable for
performing the measurements of the current, in particular when the ATLAS
version of the module is used which provides a higher measurement precision with 20 bit ADCs instead of the 16 bit ones used for the Protvino test beam runs.
\clearpage

\subsection{Results} 

The results of the run periods A and B are in very good
agreement. They clearly show that the HV current measured by the FCal1
depends linearly on the beam intensity. Combining all data a
non-linear fraction of less than \unit{0.36}{\%} at 95\% confidence level was
obtained. This applies to a beam intensity of
\unit{$1\cdot\power{10}{9}$}{p/$spill$} which corresponds to the
nominal LHC \Lumi of \unit{$10^{34}$}{\lumiunits}.\\

The result for the channel with \unit{100}{\micron} gap is
consistent with that from the \unit{250}{\micron} gap. This shows
the robustness of the method against varying gap sizes and indicates the
potential for using it during the sLHC phase as well.

\subsection{Implications for the relative luminosity measurement in ATLAS}

The differences between the fixed target experiment in Protvino and a
collider experiment like \mbox{ATLAS} have to be taken into
account when interpreting the test beam results for ATLAS. However, the proposed method is also applicable in \mbox{ATLAS} because the secondary particle flux is
proportional to the beam intensity in a fixed target experiment and to the instantaneous \Lumi in a collider experiment.\\

Furthermore, the differences between the ATLAS FCal1 and the FCALchick
have to be considered. The main differences are the different
proportions of both calorimeter modules. The tubes of the FCALchick
are nine times shorter than that of the FCal1 in ATLAS and the
FCALchick consists only of four channels with four tubes each as
described in Chapter 2. In addition, only one tube group of the
FCALchick was connected to each HV supply channel instead of 64 tube
groups in ATLAS~\cite{johnspaper}. Therefore it can
be expected that the HV current per channel and that for the whole
calorimeter will be much larger at the same particle flux in the
ATLAS FCAL1 than for the FCALchick in the test beam. Thus
statistical fluctuations per HV channel can be assumed to be further
reduced.\\

The different beam structure in Protvino compared to that at the LHC will not influence the \Lumi measurement using the FCal1 HV
currents because the low pass filters will cause an averaging over
many bunch crossings, resulting in a continuous DC current visible in
the HV current readout. It is already seen that variations of the background current and noise between the different FCal channels are present in ATLAS and also variations in time are observed. Therefore, the selection of stable channels and investigations of thresholds to separate signal from background currents are important when applying the measurement to ATLAS.

\subsection{Discussion of systematic uncertainties}

Additional systematic effects (as discussed in \cite{Bonivento:684140}) may influence the measurements. Recombination
of argon ions and electrons produced, as well as temperature effects
can be assumed as being already included in the test beam result. Furthermore, the operating conditions of the liquid argon of the calorimeter (temperature of about \unit{89}{\kelvin}, purity of about
2\,ppm) during the Protvino April 2008 run were nearly the same as expected in
ATLAS. The FCal1 HV currents caused by detector activation could be
higher than that in Protvino because of the larger detector
volume. But these currents can be neglected with respect to those due
to primary interactions in ATLAS \cite{Bonivento:684140}.\\

Therefore, only the systematic effects due to a displacement of the
mean interaction point and due to lost beam particles reaching the FCal1
could possibly deteriorate the linear response of the FCal1 HV
currents in \mbox{ATLAS} significantly. The effect of lost beam particles (e.g. due to beam-gas interactions) is
however expected to be negligible because of the forward shielding
installed in \mbox{ATLAS}~\cite{Bonivento:684140}.\\

The variations of the HV
current caused by a displacement of the mean interaction point of
\unit{1}{\centi\meter} along the beam axis is expected to be less than
0.2\,\% in each FCal1 module. This can be further reduced to 0.1\,\%
by summing over the currents of the FCal1 modules on both sides of the
\mbox{ATLAS} detector \cite{Bonivento:684140}. A possible lateral displacement of the mean interaction point would introduce a significant transverse asymmetry in the pattern of currents drawn from the HV power supplies. However, as long as such a displacement is constant in time it will not influence the linearity of the relative luminosity measurement.

\section{Summary and conclusion}

In this paper an alternative method for a relative luminosity
measurement at the LHC using the readout of the
high voltage return current in the forward section of the liquid-argon
calorimeter of \mbox{ATLAS} was discussed. The analysis of
test beam data taken at the U-70 proton accelerator in Protvino using a prototype of the forward calorimeter was
described and the proposed method was verified.\\

A linear relation between the HV current measured with the FCALchick
and the beam intensity was found with a non-linear fraction of less than
\unit{0.36}{\%} (95\% CL) at a beam intensity of \unit{1\cdot\power{10}{9}}{p/$spill$},
corresponding to the nominal \mbox{ATLAS} \Lumi of
\unit{$10^{34}$}{\lumiunits}. The most important systematic
uncertainty of approximately \unit{0.1}{\%} is expected to be caused
by variations of the mean interaction point in \mbox{ATLAS}. Taking
statistical and systematic uncertainties into account, a precision of
better than \unit{0.5}{\%} will be feasible for ATLAS.\\

\acknowledgments
The support of the IHEP staff operating the accelerator and the
beamline 23 is gratefully acknowledged. In particular we thank
V.~Anikeev, I.~Beljakov, N.~Chabrov, T.~Gurova, O.~Romashev and
I.~\mbox{Shvabovich} for their invaluable help in preparation and
running the experiment. We would also like to thank the ISEG company for their technical support.\\
This work has been supported 
by the Bundesministerium f\"ur Bildung, Wissenschaft, Forschung und Technologie, Germany, under contract number 05 HA 6OD1, 
by the German Helmholtz Alliance ''Physics at the Terascale'', 
by the Office of High Energy Physics of the U.S. Department of Energy as part of the U.S. ATLAS Research Program, 
by the Russian Fund for Basic Research and
by the Slovak Grant Agency of the Ministry of Education of the Slovak Republic and the Slovak Academy of Sciences, Project No. 2/0061/08.


\end{document}